\documentclass[a4paper]{article}

\usepackage{INTERSPEECH2019}
\usepackage{array}
\usepackage{amsmath}
\usepackage{hyperref}

\title{Large-scale Transfer Learning for Low-resource Spoken Language Understanding}

\name{Xueli Jia,  Jianzong Wang$^*$,   Zhiyong Zhang,   Ning Cheng,   Jing Xiao}
\address{Ping An Technology (Shenzhen) Co., Ltd.}

\begin{document}

\maketitle

\begin{abstract}
  End-to-end Spoken Language Understanding (SLU) models are made increasingly large and complex to achieve the state-of-the-art accuracy. However, the increased complexity of a model can also introduce high risk of over-fitting, which is a major challenge in SLU tasks due to the limitation of available data. In this paper, we propose an attention-based SLU model together with three encoder enhancement strategies to overcome data sparsity challenge. The first strategy focuses on the transfer-learning approach to improve feature extraction capability of the encoder. It is implemented by pre-training the encoder component with a quantity of Automatic Speech Recognition annotated data relying on the standard Transformer architecture and then fine-tuning the SLU model with a small amount of target labelled data. The second strategy adopts multi-task learning strategy, the SLU model integrates the speech recognition model by sharing the same underlying encoder, such that improving robustness and generalization ability. The third strategy, learning from Component Fusion (CF) idea, involves a Bidirectional Encoder Representation from Transformer (BERT) model and aims to boost the capability of the decoder with an auxiliary network. It hence reduces the risk of over-fitting and augments the ability of the underlying encoder, indirectly. Experiments on the FluentAI dataset show that cross-language transfer learning and multi-task strategies have been improved by up to $4.52\%$ and $3.89\%$ respectively, compared to the baseline. 
\end{abstract}

\let\thefootnote\relax\footnote{*Corresponding author \quad jzwang@188.com}

\section{Introduction}

Conventional SLU pipeline mainly consists of two components \cite{sutskever2014sequence}: an Automatic Speech Recognition module generates transcriptions or N-hypotheses, and a Natural Language Understanding (NLU) module classifies transcriptions into intents, in which speech recognition error propagation will be amplified during sub-sequence NLU process. Although with the rapid development of end-to-end speech recognition systems, the performance of SLU has been significant improved \cite{inaguma2019transfer,moritz2019triggered,jia2019leveraging,moritz2020streaming,miao2020transformer,inaguma2020minimum}, it still can not satisfy the application requirements, due to the complexity of scenarios.


Usually not all errors from speech recognition harm the SLU module, and those errors have no impact on the eventual performance \cite{Bhosale2019,Masumura2019}. The SLU component only keeps its attention on keywords while discarding most of the other irrelevant words \cite{Ray2018}. Thus the joint optimization approach can strengthen the focus of the model on improving the transcription accuracy that relates to target events \cite{Li2018,Gupta2018}. Recently, many efforts have been dedicated on end-to-end SLU in which the domain and the intent are predicted directly from input audio \cite{haghani2018audio,serdyuk2018towards, chen2018spoken,  renkens2018capsule,wang2020large,9053281,9053163}.
Previous researches have shown that a large amount of data is the determining factor for the excellent performance of a model \cite{serdyuk2018towards}. However, due to the lack of audio and the ambiguity of intents, it is difficult to obtain sufficient in-domain labeled data. Transfer learning methodology has become a common strategy to address insufficient of data problem \cite{xie2016transfer,huang2017transfer,tan2018survey}.
Different transfer learning strategies have been applied in SLU model and all of them result in competitive complementary results \cite{tomashenko2019investigating, caubriere2019curriculum}.
In this paper, this strategy is also applied to amplify the feature extraction capability of the encoder component, it pre-train the encoder with a large amount of speech recognition labeled data, and then transfer the encoder to the SLU model.

Recently, \cite{haghani2018audio} proposed and compared various of encoder-decoder approaches to optimize each module of SLU in end-to-end manners and have proved that intermediate text representation is crucial for SLU and jointly training the full model is advantageous. Attention-based models have been widely used in speech recognition and provide impressive performance \cite{moritz2020streaming,miao2020transformer,inaguma2020minimum,wang2019large, li2019speechtransformer, hrinchuk2019correction}.
Inspired by this, we propose a Transformer based multi-task strategy to adopt textual information in the SLU model. Since text information only acts on the decoder component in speech recognition task, it can be treated as an adaptive regularizer to adjust the encoder parameters such that contributing to improve intent prediction performance. 
It should be noticed that the lack of textual corpus is also a major challenge when training language models. To address this problem, various of methods have been carried out to expand corpus in the past decade \cite{ 222dikici2016semi, sutskever2011generating, zgank2019cross}. In addition, textual level transfer learning strategy by merging a pre-trained representation to the decoder is also explored. The pre-trained representation is obtained with the BERT model, which is designed to pre-train the deep bidirectional representations from unlabeled text by jointly conditioning on both left and right context in all layers \cite{devlin2018bert}. 

Encoder and decoder are mutual independent but are connected by the attention block, through which can get a collaborated optimization in training. 
To maximize the performance, both encoder and decoder are optimized with transfer leaning strategies. In this paper, we first propose a self-attention based end-to-end SLU structure, and applied cross-lingual transfer learning method to solve insufficient acoustic data problem. Then we propose a Transformer based multi-task strategy that conducts intent classification and speech recognition in parallel. Finally, a textual-level transfer learning structure is designed to aggregate the pre-trained BERT model into the decoder component to improves the feature extraction capability of the decoder, indirectly.

\section{Methodology}
\label{method}

In this section, a self-attention based end-to-end SLU model is proposed, Self-attention layers have been proved to be superior than recurrent layers. Next a Transformer based multi-task structure is designed to take immediate textual information into account. Finally, the CF structure is implemented in the decoder as an enhancement of the auxiliary network for the multi-task structure.

\subsection{Self-attention based End-to-end SLU}
\label{SLU_base}

Self-attention layers have been proved to be superior than recurrent layers in computational complexity when the sequence length is smaller than the representation dimensionality, and it can also yield more interpretable models than convolutional layers \cite{vaswani2017attention}.  Inspired by these advantages, an attention-based encoder-decoder structure is designed to solve SLU problems. The architecture consists of several stacks of layers. Each layer of the encoder and decoder consists of a multi-head attention module and a position-wise fully connected feed-forward network.
 A max-pooling layer is involved to aggregate the output of the encoder along time axis, as illustrated in Figure \ref{fig:general}. Softmax function is used to estimate the posterior probabilities of intents.

We denote the input acoustic frames as $ x=(x_1,...,x_T)$, where $x_t \in R^d (1 \leqslant t \leqslant T)$ indicates log-mel filter-bank (FBank) features in this work, $d$ is the dimension of Fbank, $T$ indicates the number of frames in $x$. Ground-truth posterior distribution for utterance $u$ is defined as $q^{u}=(q_1^{u},...,q_I^{u})$, which is represented as one-hot format. Cross-entropy criterion is used to evaluate the model performance, then the cost function for each utterance ${\mathcal{L}_{slu}^u}$ is defined as Equation \ref{slu_loss}.

\begin{equation} 
\label{slu_loss}
\mathcal{L}_{slu}^{u} (\theta) = -\sum_{i=1}^I q_i^{u} log p(y_i^u | x; \theta) 
\end{equation}

Where $u$ is the index of speech utterance. $\theta$ indicates model parameters. $I$ represents intent size. $y_i^u$ indicates the $i^{th}$ predicted intent and $p(y_i^u|x; \theta)$ demonstrates the posterior probability of $y_i^u$ given $x$ and $\theta$.

\subsection{Encoder Augmentation Strategies}

\subsubsection{Cross-lingual Pre-training}
Human languages share some commonality in both acoustic and phonetic aspects. Features extracted from some languages can be shared with other languages at some levels of abstraction. \cite{toth2008cross} adapted English phones on Hungarian data yield substantial gains in performance over those trained only with Hungarian data. 
Inspired by that, we concentrate on the study of cross-lingual transfer learning over the attention-based SLU model. It is achieved by pre-training the encoder with a language-specific speech signal that different from the target language. Then the encoder-decoder model is fine-tuned with a small amount of target annotation data. 

The key approach is first training a transformer-based speech recognition model with a quantity of rich resource speech and transcribed text corpus in word level, and then migrate the well-trained encoder component to the intent model. 
This can be achieved since the encoder in SLU maps the source acoustic feature to high-dimensional representation depending on large amounts of data for better representation capability, which is the same as speech recognition applications. Acoustic transfer learning make it possible to transfer representation capability of an encoder trained with rich-resource data to an intent classification task with insufficient data. In this work, we adopt the encoder from speech recognition to intent recognition directly and explore its effectiveness. 

\subsubsection{Multi-task Training}
\begin{figure}
    \centering
    \includegraphics[scale=0.35]{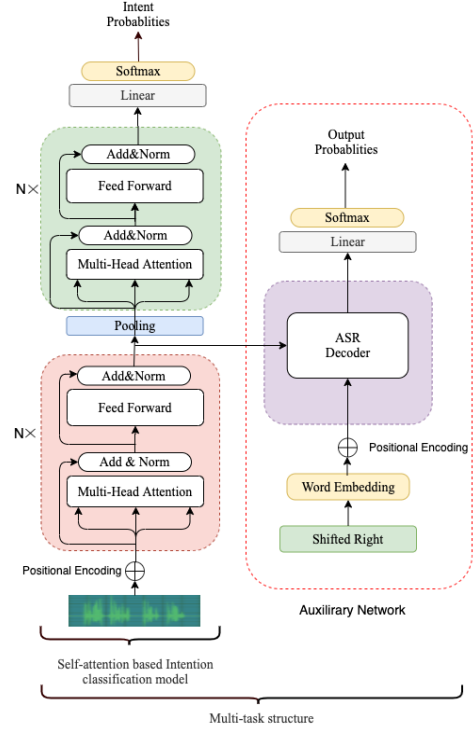}
    \caption{Structures for base model and augmentation strategies: (1) attention-based SLU model(left); (2) left encoder together with the right decoder form the basic transformer structure (3) the intent classification model together with the transformer produce the multi-task structure. }
    \label{fig:general}
\end{figure}

The multi-task structure consists of three components: an encoder module for acoustic representation, a decoder for speech recognition task, and another decoder for intent prediction task. 
The intent prediction decoder is designed to be placed after the acoustic encoder model, which is a compromised strategy compared with the conventional end-to-end SLU model, since the inaccurate prediction of text from speech recognition module. The multi-task structure is illustrated as Figure \ref{fig:general}.

In this work, intent prediction task aims at mapping the acoustic feature sequence into semantic space and treats it as semantic classification task. During this procedure, a latent operation is translating sequence of acoustic features to text, just like the task of speech recognition. So speech recognition and intent prediction have the same procedure in translating acoustic feature to high level semantic representation. Thus the multi-task architecture is designed to share the same acoustic representation for speech recognition and SLU, then optimized jointly. Since our ultimate goal is to predict intents immediately from input acoustic features. Therefore, speech recognition component can be thought as a regularizer for SLU task, and offers inductive bias to it.

\begin{figure}
    \centering
    \includegraphics[scale=0.32]{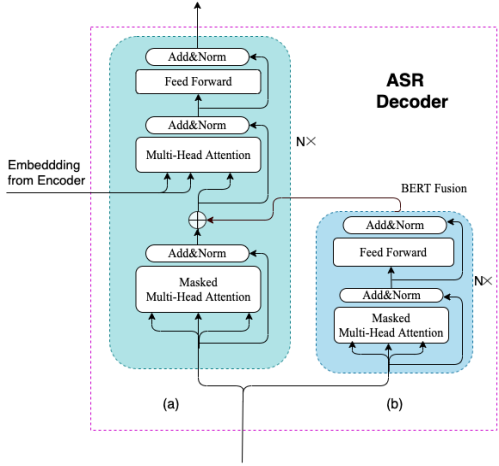}
    \caption{Structure of Speech Recognition Decoder in the auxiliary network. It consists of two sections: (a) indicates the decoder structure, it is used in both encoder pre-training and multi-task strategies; (b) is used for BERT Fusion strategy, which is to boost the decoder linguistic extraction capacity.}
    \label{fig:my_label}
\end{figure}
The same attention based model in Section 2.1 is used to do intent prediction. In order to achieve intent prediction and speech recognition in parallel, an additional stacked self-attention layers and a linear layer followed by a softmax classification layer are coupled with the encoder to output the posterior probability for speech recognition. As illustrated in Figure \ref{fig:general}, this model consists of two sub-models: an attention based intent prediction sub-model with only acoustic feature as input, and an speech recognition model accepts both the acoustic presentation from the encoder and text input from the decoder. The encoder part in the bottom left area together with the decoder component, which is detailed in Figure \ref{fig:my_label}(a), gives the typical transformer architecture. In training procedure, the loss function for speech recognition is described with cross-entropy criteria, and one-hot format is used to represent the output labels. Then the loss function for each utterance in speech recognition task can be described in Equation \ref{equ_asr_final}.

\begin{equation}
\label{equ_asr_final}
\mathcal{L}_{asr}^{u} (\theta) = \sum_{t=1}^T \mathcal{L}_{asr}^{t} (\theta) )
\end{equation}

\begin{equation}
\label{equ_asr}
\mathcal{L}_{asr}^{t} (\theta) = -\sum_{v=1}^V q_v^{t} log p(y_v^t | x; y_v ^{<t}; \theta)
\title{}
\end{equation}

Where $x=(x_1,...,x_T)$ denotes input acoustic features. $\theta$ indicates model parameters. $T$ is text length of each utterance. $V$ is vocabulary size of speech recognition. $y^t_v$ indicates the predicted token at time $t$, while $y^{<t}_{v}$ denotes the partial text sequence before $t$. Ground truth label probability distribution relates to speech recognition task $q^{t}_v=(q^1_{v},...,q^T_{v})$ is represented as one hot format. 
The loss function of the composite system is demonstrated as the combination of the SLU loss and speech recognition loss with an interpolation weight $\lambda \in [0,1]$, as shown in Equation \ref{equ_slu_asr}.

\begin{equation}
\label{equ_slu_asr}
\mathcal{L}^{u} (\theta) = \mathcal{L}_{slu}^u(\theta) + \lambda \mathcal{L}_{asr}^u(\theta)
\end{equation}

It is apparent that both SLU and speech recognition tasks have abilities of updating encoder parameters. Theoretically, we should emphasize the importance of the SLU model and intent training data since our ultimate goal is intent prediction. This is achieved by adjusting the parameter $\lambda$ to scale the effect of speech recognition. Involving speech recognition model leading to several advantages. Firstly, the quantity of annotation data in SLU task is insufficient, the encoder can produces more representative acoustic features with speech recognition training data. Then, more robust features can be extracted when it is used to compile two tasks instead of one, thus it can efficiently avoid over-fitting problem as well. 

\subsubsection{BERT Fusion}
It should be noticed that lack of in-domain text corpus is a major challenge when training language models. To address this problem, text level transfer learning strategy is explored recently. \cite{shan2019component} proposed component fusion method to incorporate externally trained neural network language model into an attention-based speech recognition system, and resulted in significant achievements. Inspired by that, we merge a pre-trained representation to the decoder to improve the performance as well.

BERT is conceptually simple and empirically powerful model and has been proved outperform many other architectures in many NLP tasks, and it is the first fine-tuning based representation model that achieves state-of-the-art performance on a large suit of sentence level and token level tasks \cite{devlin2018bert}. In this paper, we apply the BERT model to the multi-task structure to extract more powerful linguistic representation, then improves the performance of intent prediction, as shown in Figure \ref{fig:my_label}(b). 
BERT fusion only has effect on the speech recognition task, it has ability of outputting more precise text prediction. So it can be thought as an indirect way of enhancing encoder performance. The training procedure is similar with the multi-task training method, but gives text input to both the decoder and the BERT model. 

\section{Experimental setups}
\label{exp}
\subsection{Dataset}
In the experiment, two datasets are used to train and test different structures, FluentAI dataset described in \cite{lugosch2019speech} is used to train and evaluate the baseline model and SLU model with different strategies. As shown in Table \ref{tab_fluentAI}, this dataset is sampled in $16kHz$ single-channel wav format. Each audio includes a single command and is labeled with three slots: action, object, and location. 
There are $248$ different phrases with a total of $19$ hours.
The second dataset, shown in Table \ref{tab_aishell}, is a open source mandarin speech corpus AISHELL-ASR0009-OS1 which is used to pre-train the encoder component. This dataset consists of $178$ hours long speech and recorded by $400$ people from different accent areas in China. 
\begin{table}[h]
    \centering
    \caption{FluentAI Speech Command dataset}
    \begin{tabular}{c|c|c|c}
    \hline
    Split   &  Speakers & Utterances & Hours \\
    \hline
    Train & $77$ & $23,132$  & $14.7$ \\ 
    \hline
    Valid & $10$ & $3,118$ & $1.9$\\
    \hline
    Test &  $10$ & $3,793$ & $2.4$\\
    \hline
    Total & $97$ & $30,043$ & $19.0$ \\
    \hline
    \end{tabular} 
    \label{tab_fluentAI}
\end{table}

\subsection{Results and Analysis}
All experiments are conducted using 80-dim FBank feature with a frame length of $25 ms$ and a frame shift of $10 ms$. Mean and Variance normalization is applied in utterance level, then features are down-sampled by a factor of $3$, and $4$ consecutive vectors stacked at the end. 

The SLU model described in Section \ref{SLU_base} is treated as the baseline. Results in Table \ref{tab:varConfigE2ESLU} shows that the performance of SLU model have a strong correlation with the amount of parameters, this is attributing to the small amount of training dataset and complexity of the model. The best performance $91.91\%$ can be achieved when the encoder is set to $3$ layer and the decoder to $6$ layer. In subsequent experiments, these parameters are adopted to compare different enhancement strategies.

\begin{table}
    \centering
    \caption{AISHELL-ASR0009-OS1  dataset}
    \begin{tabular}{c|c|c|c}
    \hline
    Split & Hours & Male & Female \\
    \hline
    Train & $150$ & $161$  & $179$ \\ 
    \hline
    Valid & $10$ & $12$ & $28$\\
    \hline
    Test &  $5$ & $13$ & $7$\\
    \hline
    Total & $165$ & $186$ & $214$ \\
    \hline
    \end{tabular} 
    \label{tab_aishell}
\end{table}

\begin{table}
    \centering
    \caption{Results of various configures of SLU model}
    \begin{tabular}{c|c|c|c|c}
    \hline
   $N_{enc/dec}$ & $N_{head}$ & $N_{k/v}$ & $d_{m/i}$ &  Acc  \\
  \hline
    2/0 & 2 & 32 & 256/512  &  88.67 \\ 
    3/0 & 8 & 64 & 256/512  &  86.68 \\
   3/0 & 8 & 64 & 512/1024 &  90.38 \\
    \bfseries{3/6} & \bfseries{8} & \bfseries{64} & \bfseries{512/1024}   &  \bfseries{91.91} \\
    \hline
    \end{tabular}
    \label{tab:varConfigE2ESLU}
\end{table}

\begin{table}
    \centering
    \caption{Intent prediction accuracy for different strategies}
    \begin{tabular}{c|c|c|c}
    \hline
    Methodologies   &  Tune/Fix   &  Scales  &   Accuracy  \\
    \hline
    Baseline  &  - & -  & 91.91 \\ 
    \hline
    EP   &  Fix & -  &  94.86 \\
    EP   &   Fine-tune &  - &  93.25 \\
    \hline
    MT   &  - & 0.1  & 92.41 \\
    MT   &  - & 0.5  & 95.25 \\
    
    MT   &  - & 1.0  & 95.28 \\
    \hline
    MT and BF   &   Fix &  1.0 & 95.49 \\
    
    \hline
    \bfseries{EP and MT}   &   \bfseries {Fine-tune} & \bfseries{1.0 } &  \bfseries {96.07} \\
    \hline
    EP, MT and BF   &   Fine-tune, Fix & 1.0  &  94.91 \\
    \hline
    \end{tabular} \\
    \scriptsize\textit{EP: Encoder Pre-training.  FT: Fine-tune.    MT: Multi-task.   BF: BERT Fusion}
    \label{tab:sum}
\end{table}

\begin{figure}
    \centering
    \includegraphics[scale=0.25]{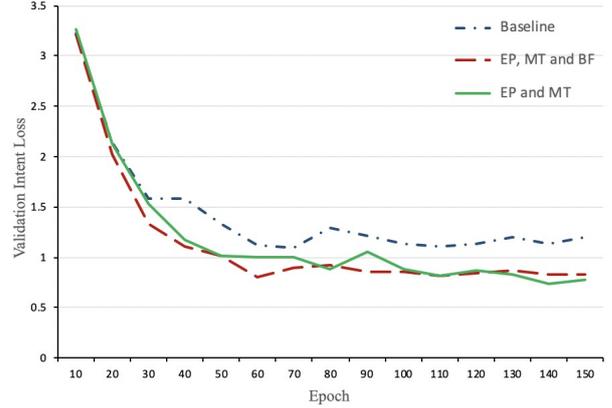}
    \caption{Validation Intent Losses for the Baseline, Multi-task with Encoder Pre-training, and Multi-task with Encoder Pre-training and BERT Fusion.}
    \label{fig:figue}
\end{figure}

Cross-lingual transfer learning is implemented by training a transformer based speech recognition model with $150$ hours of AISHELL data first, then transferring the well-trained encoder to the SLU model directly. Two experiments, fixing parameters and fine-tuning parameters of the encoder, have been conducted to check their performance. Results in Table \ref{tab:sum} indicate that both strategies have abilities of improving the performance of SLU model. It means that when training the encoder with an irrelevant language can be migrated to other language in acoustic space.
Table \ref{tab:sum} also shows that a better improvement $3.21\%$ is obtained when the encoder parameters are fixed. This implies the simplicity of the FluentAI dataset, training tends to become over-fitting when more parameters are involved. If the encoder is trained with more data, it will have more robust generation capabilities.

The multi-task experiment is implemented with FluentAI dataset as well. Table \ref{tab:sum} gives result with different speech recognition scales. It indicates that the best performance $95.28 \%$ is given when the scale is set $1.0$. It proves that the speech recognition model can bring benefits to SLU when giving a appropriate scale. Actually, it is tough to balance the parameter $\lambda$, the main point is that we want the auxiliary task to promote the shared part into two tasks in a data-driven manner, or to become a regularizer for the SLU task. The scale $1.0$ is applied in the following experiments.

BERT fusion strategy is conducted relying on the multi-task structure. The BERT
model consists of $12$ layers where each layer consists of $768$ hidden units, 12-heads, and about $110M$ parameters. Parameters of BERT model are fixed in all the subsequent experiments. Table \ref{tab:sum} indicates this strategy gives $3.58 \%$ and $0.21\%$ improvements comparing with the baseline and the multi-task method. This indicates that BERT model has capability of improving the performance of SLU model.

In addition, different combinations of these strategies are explored. Table \ref{tab:sum} demonstrates that the combination of cross-lingual pre-training and multi-task strategies obtains an accuracy of $96.07 \%$, and the combination of all these three strategies gives $94.91 \%$. Both methods produce better performance than the baseline. Figure \ref{fig:figue} depicts the validation intent loss along with epoch, both compound strategies obtain lower losses and converge faster. Theoretically, the combinations of three strategies should give the best performance. However, experiments show that 
the cross-lingual encoder pre-training with multi-task strategy gives the most positive promote on the accuracy. The reason attributes to the data sparsity, models are difficult to be well trained with limited data. And the sparsity of labeled data usually accompanies with over-ﬁtting problem, which aggravates the tuning and optimization during training procedure.

\section{Conclusion}
\label{conclusion}
In this paper, we propose an attention-based end-to-end SLU model and evaluated different augmentation strategies based on this model.  
We show that cross-lingual encoder pre-training, multi-task strategy, and BERT-fusion have abilities of improving the intent classification performance. These enhancement strategies can also extend to other areas such that improve their performance. 
Due to the limitation of data, the model is prone to over-fitting and sensitive to model parameters. More investigation on how to efficiently solve data sparsity in model training will be conducted in future. 

\section{Acknowledgement}
This paper is supported by National Key Research and Development Program of China under grant No. 2018YFB1003500, No. 2018YFB0204400 and No. 2017YFB1401202. Corresponding author is Jianzong Wang from Ping An Technology (Shenzhen) Co., Ltd.

\newpage

\bibliographystyle{IEEEtran}




\bibliography{reference.bib}

\end{document}